\documentclass[aps, epsf, amsfonts]{revtex4}
\usepackage{graphics}
\usepackage{graphicx}
\usepackage{epsfig}

\begin{document}
\input epsf.tex

\title{The Cauchy convergence of T and P-approximant templates for test-mass Kerr binary systems.}
\author{Edward K. Porter}
\affiliation{Department of Physics, Montana State University, Bozeman, MT 59719, USA.}
\vspace{1cm}
\begin{abstract}
\noindent In this work we examine the Cauchy convergence of both post-Newtonian (T-approximant) and re-summed post-Newtonian (P-approximant) templates for the case of a test-mass orbiting a Kerr black hole along a circular equatorial orbit.  The Cauchy criterion demands that the inner product between the $n$ and $n+1$ order approximation approaches unity, as we increase the order of approximation.  In previous works, it has been shown that we achieve greater fitting factors and better parameter estimation using the P-approximant templates for both Schwarzschild and Kerr black holes.  In this work, we show that the P-approximant templates also display a faster Cauchy convergence making them a superior template to the standard post-Newtonian templates.
\end{abstract}

\maketitle

\section{Introduction}
The emission of gravitational waves from double neutron stars (NS), double black holes (BH) or mixed binaries of a neutron star and a black hole, are expected to be major detectable sources for the interferometric detectors, LIGO \cite{LIGO}, VIRGO \cite{VIRGO},  GEO600 \cite{GEO}, TAMA~\cite{TAMA} and LISA~\cite{LISA}. Due to radiation reaction, the binary orbit slowly decays emitting a ``chirp'' signal (i.e. a signal whose amplitude and frequency increases with time). While there is probably a greater population of NS-NS binaries \cite{Grish,Phin,Narayan,Stairs,KalogeraAndBelczynski},  it is BH-BH binaries that are the strongest candidates for detection since they can be seen from a greater volume\cite{Grish,PostnovEtAl}.  In order to detect such sources one employs the method of matched filtering ~\cite{Helst}.  In this method, a set of theoretical waveforms or templates, that depend on a number of parameters of the source and its location and orientation relative to the detector, are cross-correlated with the detector output weighted by the inverse of the noise spectral density.  The success of matched filtering depends on how well the phase evolution of the waveform is known. To improve the recovered SNR for inspiral events, there have been many efforts to accurately compute the dynamics of a compact binary and the waveform it emits, or, to use phenomenologically defined detection template families \cite{BCV1,BCV2,BCV3}. 

Two different approximation schemes have been used to try and accurately predict the phase of the GW. On one hand, there is the post-Newtonian (PN) expansion of Einstein's equations ~\cite{BDIWW,BDI,WillWise,BIWW,Blan1,DJSABF,BFIJ} which assumes slow motion bodies.  On the other hand,  there is black hole perturbation theory \cite{Poisson1,Cutetal1,TagNak,Sasaki,TagSas,TTS} which has no restriction on velocity, but assumes that one body is much more massive than the other.  Due to the fact that the convergence of both post-Newtonian approximation and black hole perturbation theory seem to be too slow to be useful in constructing accurate templates~\cite{TTS,Cutetal2,Poisson3,Poisson4,DIS1},  Damour, Iyer and Sathyaprakash (hereafter DIS) showed for the case of a test-mass in orbit about a Schwarzschild BH, that newly constructed P-approximant templates achieved larger overlaps with the exact waveform and lower biases in the estimation of parameters than the corresponding post-Newtonian (hereafter T-approximant) templates.  While more P-approximant templates are needed to cover the same volume of parameter space~\cite{EKP}, the increased performance of the template warrants the extra computational cost.  Most recently, the DIS analysis was extended to the case of a test mass orbiting a Kerr BH \cite{portersathya}.  It was shown, once again, that the P-approximants outperformed the T-approximant templates.  For a more indepth discussion on the definition of the gravitational waveforms and the construction of the P-approximants, we refer the reader to Reference~\cite{portersathya}.

\section{The Cauchy convergence of T- and P-approximant templates.}\label{sec:cauchy}
If a sequence $\{x_{n}\}$ converges, the terms get closer and closer to the limit of the sequence as the order of the approximation increases.  The Cauchy Convergence Criterion states that {\em if and only if for every $\epsilon > 0$, there exists an integer $N$, such that for each pair of integers $m\geq N$ and $n\geq N$, $|x_{m} - x_{n}| < \epsilon$}.  The main difference between the Cauchy criterion and other convergence criteria, is that instead of demanding that the terms get closer to some limit, we demand that the terms get closer to each other.  We therefore require that for our approximations to be Cauchy convergent, as we increase the order of PN approximation we demand the condition
\begin{equation}
\left<h_{n}\left|h_{n+1}\right>\right. \rightarrow 1\,\,\,\,\,as\,\,n\rightarrow \infty,
\end{equation}
where 
\begin{equation}
\left<h\left|g\right.\right> 
=2\int_{0}^{\infty}\frac{df}{S_{h}(f)}\,\left[ \tilde{h}(f)\tilde{g}^{*}(f) +  \tilde{h}^{*}(f)\tilde{g}(f) \right],
\label{eq:scalarprod}
\end{equation}
the * denotes complex conjugate and 
$\tilde{h}(f),\, \tilde g(f)$ are the normalized Fourier transforms of $h(t),\, g(t)$.  We can see that the inverse of the detector one-sided power spectral density, $S_{h}(f)$,  acts as a weight in the definition of the scalar product.
In the case of circular equatorial orbits of a test mass around a central
black hole, the wave is parameterized by the time of arrival of the wave $t_{0}$, the phase at time of arrival $\phi_{0}$, the total mass $m$, the reduced mass ratio $\eta$ and the spin of the central black hole $q$.   In calculating the Cauchy convergence, the masses and spins of both templates are kept the same.  We therefore only need to maximize over the parameters $t_{0}$ and $\phi_{0}$.  Maximization over $t_{0}$ is achieved by simply computing the correlation of the template with the data in the frequency domain followed by the inverse Fourier transform. This yields the correlation of the signal with the data for all time-lags. 

Maximization over $\phi_{0}$ is achieved by what DIS define as the "Best Possible Overlap" when individually maximizing over the phases of two separate templates :
\begin{equation}
\left(\cos\theta_{AB}\right)_{max}=\left[\frac{A+B}{2} + \sqrt{\left(\frac{A-B}{2} \right)^{2} + C^{2}}\, \right]^{\frac{1}{2}},
\end{equation}
where
\begin{eqnarray}
A &=& \left<e_{1}^{A}\left|e_{1}^{B}\right>\right. + \left<e_{1}^{A}\left|e_{2}^{B}\right>\right.,\,\,\,\,\,\,B = \left<e_{2}^{A}\left|e_{1}^{B}\right>\right. + \left<e_{2}^{A}\left|e_{2}^{B}\right>\right., \\
C &=&  \left<e_{1}^{A}\left|e_{1}^{B}\right>\right.\left<e_{2}^{A}\left|e_{1}^{B}\right>\right. + \left<e_{1}^{A}\left|e_{2}^{B}\right>\right.\left<e_{2}^{A}\left|e_{2}^{B}\right>\right. ,\nonumber
\end{eqnarray}
and
\begin{equation}
e_{1}^{A} = \frac{\tilde{h}_{1}^{A}}{|\tilde{h}_{1}^{A}|},\ \ \ \ 
e_{2}^{A} = \frac{\tilde{h}_{2}^{A} - \left<h_{2}^{A}\left|e_{1}^{A}\right>\right.e_{1}^{A}}{|\tilde{h}_{2}^{A} - \left<h_{2}^{A}\left|e_{1}^{A}\right>\right.e_{1}^{A} |}.
\end{equation}
where $\tilde{h}_{1}^{A} = \tilde{h}\left(\phi_{0} = 0\right)$ and $\tilde{h}_{2}^{A} = \tilde{h}\left(\phi_{0} = \pi / 2\right)$.  When we calculate the Cauchy convergence, we will quote the best possible overlap in all cases.

In this paper, as we are working in the test-mass approximation, we assume that our system is composed of objects with a small mass ratio. We have thus chosen the following five systems to analyse : $(100,1.4)M_{\odot}$, $(50,1.4)M_{\odot}$, $(10,1.4)M_{\odot}$, $(100,10)M_{\odot}$ and $(50,10)M_{\odot}$.  While our results are more reliable for the more extreme mass ratio cases, it is also in our interest to examine the behaviour of templates as we approach a quasi-equal-mass case.  It was shown in reference~\cite{portersathya} that singularities plague the sub-diagonal P-approximant at the $x^{7}$ order, i.e. $P^{3}_{4}$, making this template completely useless.  Such singularities do not exist at the same order for super-diagonal P-approximants, i.e. $P^{4}_{3}$.  Because of this, when we examine the Cauchy convergence involving the $x^{7}$ approximation we will use the super-diagonal P-approximant as our template.  In this work we will use the projected third-generation EURO noise curve in our overlap calculations as it has the greatest sensitivity~\cite{sathyapsd}. 
\begin{figure}[t]
\begin{center}
\epsfig{file=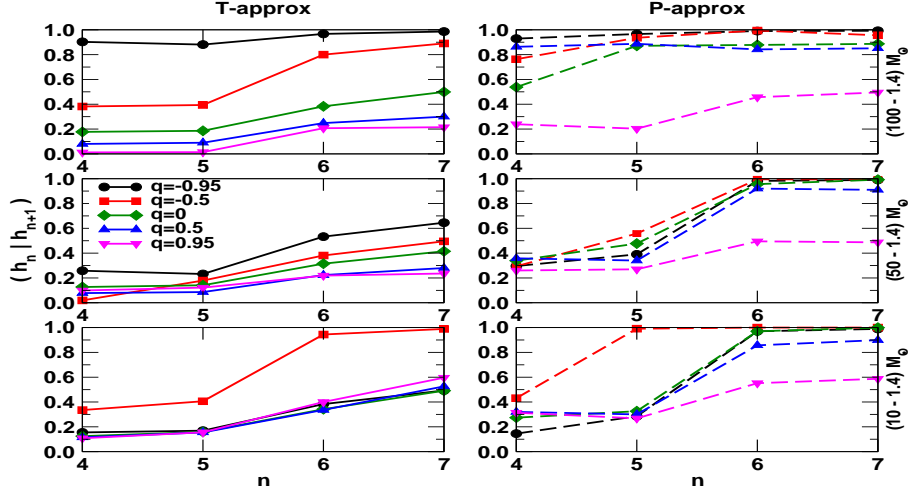, width=12cm, height=6.5cm}
\caption{The Cauchy convergence of the T and P-approximant BH-NS systems (100-1.4), (50-1.4) and (10-1.4)$M_{\odot}$ for the spin values $q=-0.95$ (circle), $q=-0.5$ (square), $q=0$ (diamond), $q=0.5$ (triangle), $q=0.95$ (upturned triangle).  The T-approximants occupy the column on the left, while the corresponding P-approximants are on the right.  All value shown are the Best-Possible-Overlaps between templates.}
\label{fig:ccbhns}
\end{center}
\end{figure}

\section{Results and Discussion.}\label{sec:results}
\subsection{Black Hole - Neutron Star.}
In Figure~(\ref{fig:ccbhns}) we have plotted the Cauchy convergence for T and P-approximants for the systems (100-1.4), (50-1.4) and (10-1.4) $M_{\odot}$.  In the high mass case we can see that the T-approximants have excellent convergence for the extreme retrograde spin at $q=-0.95$.  Once we increase the spin value and move towards prograde orbits we can see that these templates do not display good convergence properties, with the extreme prograde case at $q=0.95$ never giving an overlap of $\geq 0.2$.  As we move down the column on the left hand side of Figure~(\ref{fig:ccbhns}) we can see that the Cauchy convergence gets progressively worse as we move towards the equal mass case.  In all cases we find that the best convergence is achieved for retrograde orbits.  This should not surprise us as these are slow motion systems.  Thus, the overlaps get expectedly worse as we increase spin and decrease the total mass of the system.  In fact, in the case of a (10-1.4)$M_{\odot}$ system, we can see that besides a spin of $q=-0.5$, we obtain very bad convergence at all spin levels with the T-approximant templates.

In contrast, the P-approximant templates in the (100-1.4)$M_{\odot}$ case show excellent convergence properties.  Again we can see that the convergence begins to tail off as we increase in spin.  As we move down the cells on the right hand side of Figure~(\ref{fig:ccbhns}) toward the equal mass case, we can see that the P-approximants have slow convergence at low orders of approximation, but approach unity as we move towards the $x^{8}$ approximation.  This again is expected behaviour as we know that the P-approximants do not perform as well at lower orders of approximation.  We can however see that even in the (10-1.4)$M_{\odot}$ case we achieve overlaps of $\geq 0.9$ up to a spin of $q=0.5$.  We should remark here about the convergence properties of both templates for the extreme prograde case.  Graphically, it looks like the P-approximants perform only marginally better than the T-approximants.  However, we mentioned before that the flux function goes to zero for the T-approximants before the LSO is reached.  Thus, these templates are shorter than they should be and never reach the truly relativistic regime of the waveform.  The P-approximants on the other hand have been evolved to the LSO, so the convergence results are more reliable.  We cannot have confidence in the results from the T-approximants in the extreme prograde case, because if we could evolve these templates to the LSO we would expect the convergence to be even slower.  The P-approximants thus produce higher overlaps with a full template which point towards their superiority as templates.
\begin{figure}[t]
\begin{center}
\epsfig{file=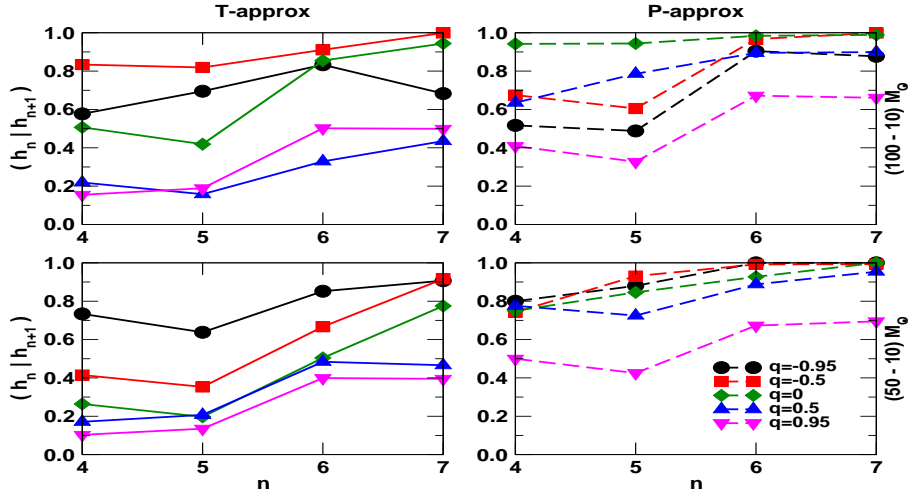, width=12cm, height=6.5cm}
\caption{The Cauchy convergence of the T and P-approximant BH-BH systems (100-10) and (50-10)$M_{\odot}$ for the spin values $q=-0.95$ (circle), $q=-0.5$ (square), $q=0$ (diamond), $q=0.5$ (triangle), $q=0.95$ (upturned triangle).  The T-approximants occupy the column on the left, while the corresponding P-approximants are on the right.  All value shown are the Best-Possible-Overlaps between templates.}
\label{fig:ccbhbh}
\end{center}
\end{figure}
\subsection{Black Hole - Black Hole.}
In Figure~(\ref{fig:ccbhbh}) we have plotted the convergence results for the systems (100-10) and (50-10)$M_{\odot}$.  If we again focus on the cells in the left hand column, we can see that we do not have the same pattern of performance for the T-approximant templates.  For the (100-10)$M_{\odot}$ case, we can see the T-approximants obtain the fastest convergence as we move towards the Schwarzschild case.  Once we increase the spin of the central BH and move onto prograde orbits, we again see a drop-off in the convergence rate.  In the (50-10)$M_{\odot}$ case we can see that the Cauchy convergence gets progressively worse as we move from retrograde to prograde motion.  The best overlaps are again obtained from extreme retrograde orbits, while the worst are from the extreme prograde.

Once again, focusing on the cells in the right hand column, we can see that the P-approximants have better convergence properties.  While in the retrograde case for the (100-10)$M_{\odot}$ the convergence is slower, we can see that for both systems the overlaps approach unity as we move to higher approximation.  In most cases, while the convergence is slower at lower orders of approximation, it is still accelerated in comparison to the T-approximant templates.  Again we can see that while displaying a faster convergence, the P-approximants at $q=0.95$ still have a poor Cauchy convergence.  While it is obvious that the P-approximants are superior templates to search for waveforms up to a spin of $q\sim 0.5$ (and possibly slightly higher), they may still not be the best templates to search for extreme prograde systems. 


\section{Conclusions}\label{sec:conclusions}
We have looked at the Cauchy convergence of both T and P-approximant templates in the test-mass approximation for Kerr binaries.  We examined the BH-NS and BH-BH cases for both prograde and retrograde orbits using the third generation detector, EURO.  In a previous work it was shown that P-approximant templates achieve higher fitting factors with lower biases in the estimation of parameters compared to the T-approximants.  In this study we have shown that in all examined cases with various spin values, the P-approximants have a faster Cauchy convergence than the T-approximant templates.  We found that the T-approximants show good convergence for extreme mass-ratio retrograde motion.  However in all cases there is a serious retardation in convergence as we move to Schwarzschild and prograde orbits.  We also see a degradation in overlaps as we move towards the equal mass case.  The P-approximants, while slow to converge at low orders of approximation, approach unity in all cases as we increase the approximation order.  This is due to the known fact that the P-approximants perform better at higher orders of approximation.

While convergence is slow in both cases for the extreme prograde case, we have more faith in the P-approximants due to the fact that we can evolve these templates to the LSO.  This isn't possible with the T-approximants as the flux function goes to zero before the LSO is reached.  In fact, these templates should not be used to search for systems with a spin of $q\geq 0.5$.  This work reinforces the conclusion of an earlier work that P-approximant templates are more reliable than T-approximants.  They give better fitting factors, have lower biases in the estimation of parameters, and as shown in this work, display an accelerated Cauchy convergence.   Due to these results, these templates should be used in any data-analysis strategy to detect gravitational waves from binary systems.

\section*{Acknowledgments}
The author would like to thank Prof. Sathyaprakash for suggesting the problem and for his useful comments.
\\

\clearpage

\end{document}